\documentclass[final,1p,times,authoryear]{elsarticle}

\usepackage{amssymb}
\usepackage{amsmath}
\usepackage{hyperref}

\newcommand{\exodata}{ExoData}

\journal{Computer Physics Communications}

\begin{document}

\begin{frontmatter}

\title{ExoData: A Python package to handle large exoplanet catalogue data}

\author{Ryan Varley}

\address{Department of Physics \& Astronomy, University College London\\
       132 Hampstead Road, London, NW1 2PS, United Kingdom\\
              r.varley@ucl.ac.uk  }

\begin{abstract}
Exoplanet science often involves using the system parameters of real exoplanets for tasks such as simulations, fitting routines, and target selection for proposals. Several exoplanet catalogues are already well established but often lack a version history and code friendly interfaces. Software that bridges the barrier between the catalogues and code enables users to improve the specific repeatability of results by facilitating the retrieval of exact system parameters used in an articles results along with unifying the equations and software used. As exoplanet science moves towards large data, gone are the days where researchers can recall the current population from memory. An interface able to query the population now becomes invaluable for target selection and population analysis.

\exodata{} is a Python interface and exploratory analysis tool for the Open Exoplanet Catalogue. It allows the loading of exoplanet systems into Python as objects (Planet, Star, Binary etc) from which common orbital and system equations can be calculated and measured parameters retrieved. This allows researchers to use tested code of the common equations they require (with units) and provides a large science input catalogue of planets for easy plotting and use in research. Advanced querying of targets are possible using the database and Python programming language. ExoData is also able to parse spectral types and fill in missing parameters according to programmable specifications and equations. Examples of use cases are integration of equations into data reduction pipelines, selecting planets for observing proposals and as an input catalogue to large scale simulation and analysis of planets.

\exodata{} is a Python package available freely on \textit{GitHub}\footnote{\url{https://github.com/ryanvarley/exodata}}. It is open source and community contributions are encouraged. The package can be easily installed using \textit{pip install exodata}, detailed setup information is provided within.
\end{abstract}

\end{frontmatter}

{\bf PROGRAM SUMMARY}

\begin{small}
\noindent
{\em Manuscript Title:} ExoData: A Python package to handle large exoplanet catalogue data                                       \\
{\em Authors:} Ryan Varley                                              \\
{\em Program Title:} ExoData                                     \\
{\em Journal Reference:}                                      \\
{\em Catalogue identifier:}                                   \\
{\em Licensing provisions:} GPL 3.0                                   \\
{\em Programming language:} Python 2.7, 3.4, 3.5                                   \\
{\em Computer:} Any                                             \\
{\em Operating system:} Any                                     \\
{\em RAM:} less than 200MB                                              
{\em Classification:} 1.7	Stars and Stellar Systems                           \\
{\em External routines/libraries:} numpy, quantities, matplotlib, requests, astropy, seaborn                            \\
{\em Nature of problem:}\\
  Being able to use exoplanet catalogue values in code including where there may be incomplete and incorrectly formatted values. Also being able to use the whole catalogue data at once, both for user querying, visualisation and in large simulation programs.
   \\
{\em Solution method:}\\
  An interface to access the catalogue including filling in missing values and parsing of the catalogue data. Creating an API usable by both humans and other code, implementation of commonly used exoplanet equations, a plotting library.
   \\
{\em Running time:}\\
  A few seconds depending on the task.
   \\
\end{small}

\section{Introduction}
\label{sec:intro}

The field of exoplanets is rapidly expanding with current population now in the thousands and new measurements published increasingly frequently. A catalogue of Exoplanets is necessary to keep track of these systems and their parameters, ideally being open and editable by all with a version history to enable researchers to reproduce results using the exact same values. This is especially important in large scale simulations and other work on multiple targets where a catalogue version may be more easily given over individual values and sources.

A second obstacle lies in the interface between the catalogue and the code. Such an interface can add value in its ability to calculate values using published equations, easily generate plots, estimate parameters, while keeping the catalogue up to date. It can also take into account all the fringe cases which can trip up a standard loading code (such as certain missing values and planets without a host star). The ability to replicate the catalogue is further enhanced through quoting both a catalogue and interface version. The exact parameters used for every target in the catalogue can then be easily obtained including any calculated parameters and estimations by \exodata. This tool then allows instant access to the exact parameters used in a paper.

\subsection{Open Exoplanet Catalogue}

The Open Exoplanet Catalogue \citep[OEC,][]{Rein2012} is an open-source, version controlled catalogue of exoplanets and their system parameters. It is hierarchical, preserving the format of the system (including binary star layouts). OEC uses \textit{git} for version control and the catalogue exists as a series of XML files (one per system). This format makes OEC much more diverse than most other exoplanet catalogues by letting users create their own `forks' of the catalogue where they can make their own changes and adaptations whilst still able to receive updates from the original. The advantages of version control include being able to see the exact changes made to each version, easily roll back the catalogue to any previous state and report the exact version of the catalogue your using to other researchers (using the commit SHA-1\footnote{In \textit{git} versions each commit (snapshot) of the code has a SHA-1 hash of the source which is functionally unique and is used to reference that version of the code.}).

The open nature of the catalogue means that anyone can download the full database and history of the catalogue, contribute their own changes and set up their own branched versions.

\section{Exoplanet Background Theory}\label{sec:bg-theory}

An exoplanet is a planet that orbits a star other than the Sun. We can describe a simple exoplanet system with a planet orbiting a star with a stellar radius of $R_\star$ stellar mass of $M_\star$ planetary radius $R_p$, planetary mass $M_p$ and semi-major axis of the orbit $a$.

\subsection{Describing the star}

The stellar luminosity $L_\star$ is given by the Stefan-Boltzmann equation applied to the surface area of a sphere,
\begin{equation}
\label{eqn:luminosity}
L_\star = 4\pi R^2_\star \sigma T^4_\star
\end{equation}
where $\sigma$ is the Stefan-Boltzmann constant. Stellar temperature can be estimated using the stellar mass-radius relation described in \citet{Cox2000}:
\begin{equation}
\label{eqn:stellar-r}
R_\star = kM^x_\star
\end{equation}

\noindent where k and x are constants for each stellar sequence. We can also describe stellar temperature using the main sequence relationship
\begin{equation}
\label{eqn:stellar-t}
T_\star = 5800 \times M_\star^{0.65}.
\end{equation}

The stars luminosity distance $d$ (with negligible extinction) is given by rearranging the absolute magnitude relation
\begin{equation}
\label{eqn:star-distance}
m-M = 5 \log_{10}{d} -5
\end{equation}

\noindent where $m$ is the apparent magnitude and $M$ is the absolute magnitude. Whilst $m$ is easily measured and commonly known $M$ is often undefined. We can estimate $M$ for a star using an absolute magnitude lookup table\footnote{\url{http://xoomer.virgilio.it/hrtrace/Sk.htm_SK3} from Schmidt-Kaler (1982)} based on spectral type.

In cases where we do not have a measured stellar magnitude in the band required but do have a measured value for another band we can use the conversion factors given in table A5 of \citet{Kenyon1995} to convert between magnitudes based on the stellar spectral type.

\subsection{Describing the planet}

We can infer the period of the orbit $P$ using Kepler's third law
\begin{equation}\label{eqn:keplersthird}
P = \sqrt{\frac{4\pi^2a^3}{G \left(M_\star + M_p \right)}}
\end{equation}

\noindent where $G$ is the gravitational constant. We can rewrite this in terms of the semi-major axis given the stellar mass and period. Note that the $M_p$ term is often excluded as $M_\star \gg M_p$.
\begin{equation}
\label{eqn:sma}
a = \left( \frac{P^2 G M_*}{4*\pi^2} \right)^{1/3}
\end{equation}

The semi-major axis can additionally be inferred from the temperature of the planet $T_p$ and host star $T_\star$ given an albedo for the planet $A_p$ and a greenhouse constant $\epsilon$ \citep{Tessenyi2012}:
\begin{equation}\label{eqn:greenhouse}
a = \sqrt{\frac{1-A_p}{\epsilon}} \frac{R_\star}{2} \left(\frac{T_\star}{T_p}\right)^2 .
\end{equation}

The mean planetary effective temperature can be expressed by evaluating the radiation in and out of the planet with a greenhouse effect contribution $\epsilon$ using a rearranged version of Eqn. \ref{eqn:greenhouse}
\begin{equation}
\label{eqn:planet-temp}
T_{pl} = T_\star \left( \sqrt{\frac{(1-A)}{\epsilon}\frac{R_\star}{2a}}\right)^{1/2}.
\end{equation}

From Newton's law of Gravitation we can calculate the surface gravity of the planet
\begin{equation}
\label{eqn:surface-grav}
g = \frac{GM}{R_p^2}.
\end{equation}

This can be expressed as $\log{g}$, the base 10 $\log$ of $g$ in CGS units.

If the mass of a planet  is unknown, according to mass-radius relationships of known exoplanets (e.g. \citet{Grasset2009}) the mass can be crudely estimated for a given planetary radius by assuming the density of the planetary class it is likely to be in. For this purpose we infer super-Earths as Earth density, and Jupiter and Neptune like planets with their respective densities. Assumed values are described later in Table \ref{tab:type-assum} but are easily programmable (see assumptions module section \ref{sec:assumptions}).

The scale height $H$ is the increase in altitude for which the atmospheric pressure decreases by a factor of e.
\begin{equation}
\label{eqn:scale-height}
H = \frac{k T_p}{\mu g}
\end{equation}

\noindent where $\mu$ is the mean molecular weight of the planetary atmosphere and $g$ is the planetary surface gravity.

\subsection{Transiting Exoplanets}

An exoplanet transits if it can be observed passing in front of its host star. During a transit the planet occults light from the star with the change in flux at mid transit known as the transit depth. The transit depth can be estimated as the square of the ratio of the planetary radius to the stellar radius (however, it is possible for a planet to graze the limb creating a partial transit).

The impact factor $b$ is the projected distance between the planet and star centres during mid transit and it is described as \citep{Seager2003}:

\begin{equation}
\label{eqn:impact-parameter}
b \equiv \frac{a}{R_*} \cos{i}.
\end{equation}

If the orbit is circular, we can calculate the duration of the transit $T_\text{14}$ from first to last contact using Eqn. 3 from \citep{Seager2003}:

\begin{equation}
\label{eqn:duration}
T_\text{14} = \frac{P}{\pi}\arcsin \left(\frac{R_\star}{a}\frac{\sqrt{(1+k)^2 + b^2}}{\sin{a}} \right)
\end{equation}

\noindent where k is $\frac{R_p}{R_s}$. If the planet has an eccentric orbit where the exact solution is more complex we adopt the approximation described by \citet{Kipping2011}:

\begin{equation}
\label{eqn:duration-kipping}
T_{14} = \frac{P}{\pi} \frac{\varrho^2_c}{\sqrt{1-e^2}} \arcsin{\left( \frac{\sqrt{1-a_R^2\varrho_c^2\cos^2{i}}}{a_R\varrho_c\sin{i}} \right)}
\end{equation}

\noindent where $a_R = (a / R_\star)$ and $\varrho_c$ is the planet-star separation at the moment of mid-transit in units of stellar radii

\begin{equation}
\varrho_{P,T}(f_p) = \frac{1-e^2_P}{1+e_P \sin(\omega)}
\end{equation}

\noindent where $\omega$ is the argument of pericentre and $b_{P,T}$ is the adjusted impact parameter described by

\begin{equation}
b_{P,T} = (a_P / R_\star)\varrho_{P,T}\cos{i}.
\end{equation}

\section{Installation}

\exodata{ }is a Python package tested on Python versions 2.7, 3.4 and 3.5. The easiest way to get the package is through the Python packaging tool \textit{pip},

\begin{verbatim}
pip install exodata
\end{verbatim}

Alternatively the source code can be downloaded from \url{https://github.com/ryanvarley/exodata}. The package can then be installed by the following from the command line in the package directory.

\begin{verbatim}
python setup.py install
\end{verbatim}

The package has a test suite which can be used to verify the package is working as expected. To run the tests use

\begin{verbatim}
python setup.py test
\end{verbatim}

You can choose to use the package without downloading your own copy of the catalogue, instead choosing to automatically fetch the latest version each time. If you want to use your own version you will need to download a version of the Open Exoplanet Catalogue (OEC)\footnote{\url{https://github.com/OpenExoplanetCatalogue/open_exoplanet_catalogue}} to your machine. The easiest way to do this and keep the catalogue up to date is through \textit{git}. Move to the folder you want to store the catalogue in and then

\begin{verbatim}
git clone https://github.com/OpenExoplanetCatalogue/open_exoplanet_catalogue.git
\end{verbatim}

You can then `import exodata' in Python and setup the catalogue.
\begin{verbatim}
import exodata
databaseLocation = "/open-exoplanet-catalogue/systems/"
exocat = exodata.OECDatabase(databaseLocation)
\end{verbatim}

\noindent where \textit{databaseLocation} should be the full path to the systems folder in the OEC directory or any other folder that contains OEC style XML files you want to use.

To update the catalogue move to the folder where you downloaded the catalogue and type
\begin{verbatim}
git pull origin master
\end{verbatim}

When using \exodata{ }in publications you should give the commit SHA-1 of the OEC version used and the \exodata{} version number. This document was created with the Open Exoplanet Catalogue (dc8c08a4ba0c64dd039e96c801d12f17c82a7ff3, 1st May 2016) using \exodata{} Version 2.1.5.

\subsection{Dependencies}

The following dependencies are required to run \exodata. They are installed automatically by \textit{setuptools} if required when following the above installation procedure.

\begin{itemize}
\item numpy
\item matplotlib
\item quantities (Product Quantities)
\item requests
\item seaborn
\end{itemize}

\section{Usage}

\exodata{ }is split into a series of modules dealing with the exoplanet database, equations, plots and units (see Table \ref{tab:etlos-modules} for a list with descriptions).

\begin{table}
    \begin{tabular}{lp{12cm}}
    Module          & Descrption                                                                                                                                                                          \\ \hline
    Assumptions     & Holds classification assumptions such as at what mass or radius a planet is defined as a super-Earth.                                                                                \\
    Astroclasses    & Classes for the System, Binary, Star and Planet object types.                                                                                                                        \\
    Astroquantities & Expands the \textit{Product Quantities} Python package with astronomical units like Solar Radius and compound units such as $g/cm^3$.                                                         \\
    Database        & Holds the database class and the various search methods.                                                                                                                             \\
    Equations       & Implementation of exoplanet related equations including orbital equations, planet and star characterisations and estimations. \\
    Example         & Generate example systems for testing code.                                                                                                                                           \\
    Flags           & Each object has a flag object attached which lets you know which assumptions have been made such as `calculated temperature'.                                                        \\
    Plots           & Plot functions for common plot types that can be used to to easily display data from the catalogue.                                                                                  \\
    \end{tabular}
    \caption{Table showing the modules available in \exodata{ }and what they are used for.}
    \label{tab:etlos-modules}
\end{table}

The code contains 5 main objects types, the database which holds all objects and provides functionality (such as searching) and the astronomical objects (Systems, Binaries, Stars and Planets). Moons and other types can be easily added when needed.

Like the Open Exoplanet Catalogue (OEC), the full structure of a system is preserved. Planets are children of the star (or binary) they orbit, stars are grouped in binaries where present and binaries can exist within a binary. This offers a significant structural advantage over most linear formats.

Note that the examples that follow in this paper display the raw output from the console which means most numbers are displayed showing 11 decimal places which include any floating point errors and so should not be taken as the true uncertainty in the measurement.

\subsection{As a catalogue interface}

The database is initialised by providing the OECDatabase object with the catalogue location on your machine.

\begin{verbatim}
import exodata

# Either using your own catalogue version
databaseLocation = "/open_exoplanet_catalogue/systems/"
exocat = exodata.OECDatabase(databaseLocation)

# OR downloading the latest version each time
exocat = exodata.load_db_from_url()
\end{verbatim}
The \textit{exocat} variable now contains the initialised database which can be accessed in several ways. There are lists for each type of object (i.e. \textit{exocat.planets}, \textit{exocat.stars} etc), a dictionary for each object (i.e. \textit{exocat.planetDict}) and a list of transiting planets (\textit{exocat.transitingPlanets}).

You can retrieve a particular planet either by searching or giving the exact name.

\begin{verbatim}
>>> exocat.planetDict["Gliese 1214 b"]
Planet("Gliese 1214 b")

>>> exocat.searchPlanet("gj1214b")
Planet("Gliese 1214 b")
\end{verbatim}

We can then traverse the hierarchy for the planet.

\begin{verbatim}
>>> kepler10b = exocat.planetDict["Kepler-10 b"]
>>> kepler10b.parent
Star("Kepler-10")

>>> kepler10 = kepler10b.star
>>> kepler10.planets
Planet("Kepler-10 b"), Planet("Kepler-10 c")]

>>> kepler10b.system
System("Kepler-10")
\end{verbatim}
\subsubsection{Retrieving Values}

With an object selected we can query information from the database.

\begin{verbatim}
>>> gj1214b = exocat.planetDict["Gliese 1214 b"]
>>> print gj1214b.R  # planetary radius
array(0.2525) * R_j

>>> print gj1214b.a  # orbit semi-major axis
array(0.01488) * au

>>> print gj1214b.star.M  # stellar mass of host star
array(0.176) * M_s
\end{verbatim}
The returned values are \textit{Product Quantities} objects that act like floats or more precisely a single value \textit{numpy} array but with a few extra features such as unit rescaling and dimensionality checking (see section \ref{sec:units}).

\begin{verbatim}
>>> import exodata.astroquantities as aq
>>> print gj1214b.R.rescale(aq.R_e)  # rescale to earth radii
array(2.770762439177523) * R_e

>>> print gj1214b.M / gj1214b.R  # mass / radius ratio
array(0.07722772277227723) * M_j/R_j
\end{verbatim}
In addition to retrieving values straight from the database we can also directly call some equations from the \textit{exodata.equations} module with the variables pre-filled (see \ref{sec:methods-per-object} for a full list).

\begin{verbatim}
>>> print gj1214b.calcDensity()
array(1.6067887724162309) * g/cm**3

>>> print gj1214b.calcTemperature()
array(606.022512904359) * K
\end{verbatim}

\subsubsection{Querying the Database}
Querying can be performed on object list in the form of Python \textit{for} loops or list comprehensions. \exodata{ }will return \textit{numpy.nan} if a value is absent to avoid breaking loops and comparisons with \textit{exodata.MissingValue} exceptions. MissingValue Exceptions will still be recorded in the logfile and so can be examined if needed. An example of such a query is returning the planets discovered using the radial velocity method.

\begin{verbatim}
>>> detected_by_rv = [planet for planet in exocat.planets
                  if planet.discoveryMethod =="RV"]
>>> print len(detected_by_rv) # number of planets detected by RV
658

>>> print detected_by_rv[:2]  # first 2 planets detected by RV (alphabetical)
[Planet("11 Com b"), Planet("11 UMi b")]
\end{verbatim}
You can use this same method to get a list of every planets radius. Note that some values will return as \textit{numpy.nan}, we can filter these in the initial query so a normal \textit{numpy.mean} will work.

\begin{verbatim}
>>> import numpy as np
>>> radius_list = [planet.R for planet in exocat.planets
               if planet.R is not np.nan]
>>> print np.mean(radius_list)
0.454992066547
\end{verbatim}
More advanced queries with additional \textit{if} clauses can be used to filter the results further. For example, to display all planets with a planetary radius between 1.83 and 1.9 Jupiter radii.

\begin{verbatim}
>>> print [planet for planet in exocat.planets
        if planet.R is not np.nan
        and 1.83*aq.R_j < planet.R < 1.9*aq.R_j]
[Planet("KELT-8 b"), Planet("TrES-4 A b")]
\end{verbatim}
For complicated queries a Python \textit{for} loop may be clearer. Here we show all planets with a radius between 1.5 and 1.7 Jupiter radii that were discovered by the transit method and have an orbital eccentricity greater than 0.1.

\begin{verbatim}
>>> planet_list = []
>>> for planet in exocat.planets:
        if 1.5*aq.R_j < planet.R < 1.7*aq.R_j:
            if planet.discoveryMethod == "transit":
                if planet.e > 0.1: # eccentric
                    planet_list.append(planet)
>>> print planet_list
[Planet("HATS-3 b"), Planet("Kepler-447 b"), Planet("WASP-90 b")]
\end{verbatim}

\subsection{Units}\label{sec:units}
In exodata, all equations use and require units. This means values can easily be rescaled and all equations are dimensionally checked. Units are provided through the \textit{product quantities} package\footnote{See \url{http://pythonhosted.org/quantities/}.} which provides all the standard units and constants provided by the National Institute of Standards and Technology (NIST). Some astronomical units (see table \ref{tab:units} for the list of values and sources) are added through the internal \textit{exodata.astroquantities} module which imports all \textit{Product Quantities} units for ease of use. We can then import the units from both modules using:

\begin{verbatim}
>>> import exodata.astroquantites as aq
\end{verbatim}
\textit{Product Quantities} adds units to numbers as a special numpy array type. A unit is added to a value by multiplying the values by the unit i.e.

\begin{verbatim}
>>> planetRadius = 10 * aq.R_j
\end{verbatim}
\textit{Product Quantities} provides several methods for dealing with values with a unit. Rescaling can be done using the \textit{.rescale} method e.g. to rescale solar mass to units of Earth's mass.

\begin{verbatim}
>>> print (1.99e+30 * aq.kg).rescale(aq.M_e)
array(333211.10011570295) * M_e
\end{verbatim}
Units can be converted to their unit-less counterparts (i.e for passing to code without unit support) by wrapping them in a \textit{float()} or \textit{np.array()} argument. You should always remove units with a rescale to ensure the value is in the unit you expect.

\begin{verbatim}
>>> value = 100 * pq.cm
>>> print float(x.rescale(pq.m))
1.0
\end{verbatim}

\subsection{Equations}

The equations module implements the equations described in section \ref{sec:bg-theory} as Python classes. The classes take input of every variable bar one and will output the variable left out. They therefore implement all permutations of the equation with respect to the variables.

As an example we choose the equation Kepler's Third Law (Eqn. \ref{eqn:keplersthird}), a full list of implemented equation classes and the equation they draw from are given in \ref{sec:equation-table}. The function is called \textit{KeplersThirdLaw} and taking the values for GJ 1214 b we can calculate the period.

\begin{verbatim}
>>> from exodata.equations import KeplersThirdLaw
>>> KeplersThirdLaw(a=0.01488*aq.au, M_s=0.176*aq.M_s).P
array(1.579696141940911) * d
\end{verbatim}

Similarly, given the other two values we can calculate the missing one.

\begin{verbatim}
>>> KeplersThirdLaw(a=0.015*aq.au, P=1.58*aq.d).M_s
array(0.18022315673929146) * M_s

>>> KeplersThirdLaw(M_s=0.176*aq.M_s, P=1.58*aq.d).a
array(0.01488190807285067) * au
\end{verbatim}

\subsection{Assumptions}\label{sec:assumptions}

The assumptions module handles how non universally defined parameters are set along with how some missing values may be filled. This include how planets are categorised (i.e. super-Earth) and allows researchers to use their own definitions for these planet classes along with adding new classes.

By default \exodata{ }sets these to commonly assumed values. Table \ref{tab:type-assum} shows how a planet is classified as being a Jupiter, Neptune or super-Earth along with the assumptions of mean molecular weight and density where required. For the mass limits we use the classification boundaries adopted by \citep{Tinetti2013}.

\begin{table}
    \begin{tabular}{l|lll}
    ~                            & Jupiter & Neptune & Super-Earth \\ \hline
    Mass ($M_\oplus$)            & $>$ 20  & $>$ 10  & $<$ 10      \\
    Radius ($R_\oplus$)        & $>$ 6   & $>$ 4   & $<$ 1.8       \\
    Density (g/$cm^3$)           & 1.3   & 1.6   & 4           \\
    Mean Molecular Weight (a.m.u) & 2.0       & 2.3       & 18          \\
    \end{tabular}
    \caption{Default assumptions of planet values per classification. We use the mass to classify a planet first, if  the mass is missing we then use the radius. Densities are Jupiter, Neptune and Earth densities. All these values can be changed in the program (see text).}
    \label{tab:type-assum}
\end{table}

Assumptions are implemented to make it easy to change these limits and also add more by editing the assumptions dictionary located at \textit{exodata.assumptions.planetAssumptions}.

This dictionary has the keys `masstype', `radiusType' and `tempType' which define how a planet is classified. Each key is a list of rules defining a separate type of planet in the format \textit{(upperlimit, name)}. For example we have defined the following for `massType'

\begin{verbatim}
[(10 * aq.M_e, "Super-Earth"),
 (20 * aq.M_e, "Neptune"),
 (float("inf"), "Jupiter")]
\end{verbatim}
Setting the last rule to infinity defines the last rule having no upper limit. The first rule is assumed to have no lower limit. The rules should be listed in descending order of magnitude and can be added to, modified or removed as necessary.

The dictionary also has keys defining certain value assumptions based upon the classifications set previously. These are `mu', `albedo' and `density'. These are fed into the planet classes and can be used in calculations if needed. Instead of containing a list they contain a dictionary which takes a classification `name' defined previously and provides a value for that class. Both the `mu' and `density' keys are defined by mass and radius types, for example:

\begin{verbatim}
{"Super-Earth": 18 * aq.atomic_mass_unit,
 "Neptune"    : 2.3 * aq.atomic_mass_unit,
 "Jupiter"    : 2 * aq.atomic_mass_unit}
\end{verbatim}
Again these values can be changed and modified as needed and new rules added for any new classes added to the previous set of rules.

\subsection{Flags}\label{sec:flags}

When a catalogue value is missing it is calculated or estimated using an appropriate equation in \textit{exodata.equations} if possible. While this is in many cases desirable some assumptions are more accurate than others and it is useful to know when a parameter has been filled. To this end we raise a flag in an object if a certain actions have occurred. The list of flags and the functions that raise them is shown in Table \ref{tab:flags}.

\begin{table}
\caption{List of flags and the functions that raise them when a parameter is missing and is calculated or estimated instead.}
    \begin{tabular}{ll}
    Flag                   & Calculated using \\
    \hline
    Estimated Mass         & estimateDistance()   \\
    Calculated SMA         & calcSMA()         \\
    Fake                   & Any planet that is in the `Fake Planets' list in the xml files         \\
    Estimated Distance     & estimateDistance()        \\
    Estimated magV         & magV (V magnitude)         \\
    Calculated Period      & calcPeriod()      \\
    Calculated Temperature & calcTemperature()         \\
    \end{tabular}
    \label{tab:flags}
\end{table}

For example when we ask for the semi-major axis of a planet (\textit{planet.a}) and a value is not recorded in the catalogue the interface will attempt to calculate it using Kepler's Third Law. If it is successful the value will be returned and the flag `Calculated SMA' will be raised. A list of flags that have been raised in an object can be seen by looking at the flags object of planet or star class (i.e \textit{planet.flags}).

If instead you want only raw catalogue values with no attempt at filling in missing values you can turn off this behaviour using by setting \textit{params.estimateMissingValues=False}.

\subsection{Plotting with \exodata}

\exodata{ }includes some graphing functions for easily displaying catalogue data. The main types of plots available are parameter against parameter and the number of planets per parameter bin. The general format is setting up the plot class with a list of objects to plot and then calling a plot function of the type you want with any other visual arguments. The plot style can be changed using standard \textit{matplotlib} styles. In our examples we used the `whitegrid' style by the Seaborn module\footnote{See \url{http://stanford.edu/~mwaskom/software/seaborn/} for more information}. In order for plots to be visible in the command line you may need to import \textit{matplotlib}\footnote{import matplotlib.pyplot as plt} and use the \textit{plt.show()} after calling a plot function.

\subsubsection{Number of planets by discovery method}

\begin{figure}[htbp]
  \caption{Example plot that can be generated with \textit{DiscoveryMethodByYear} in the \exodata{ }plotting module. See text for description.}
  \centering
    \includegraphics[width=0.7\textwidth]{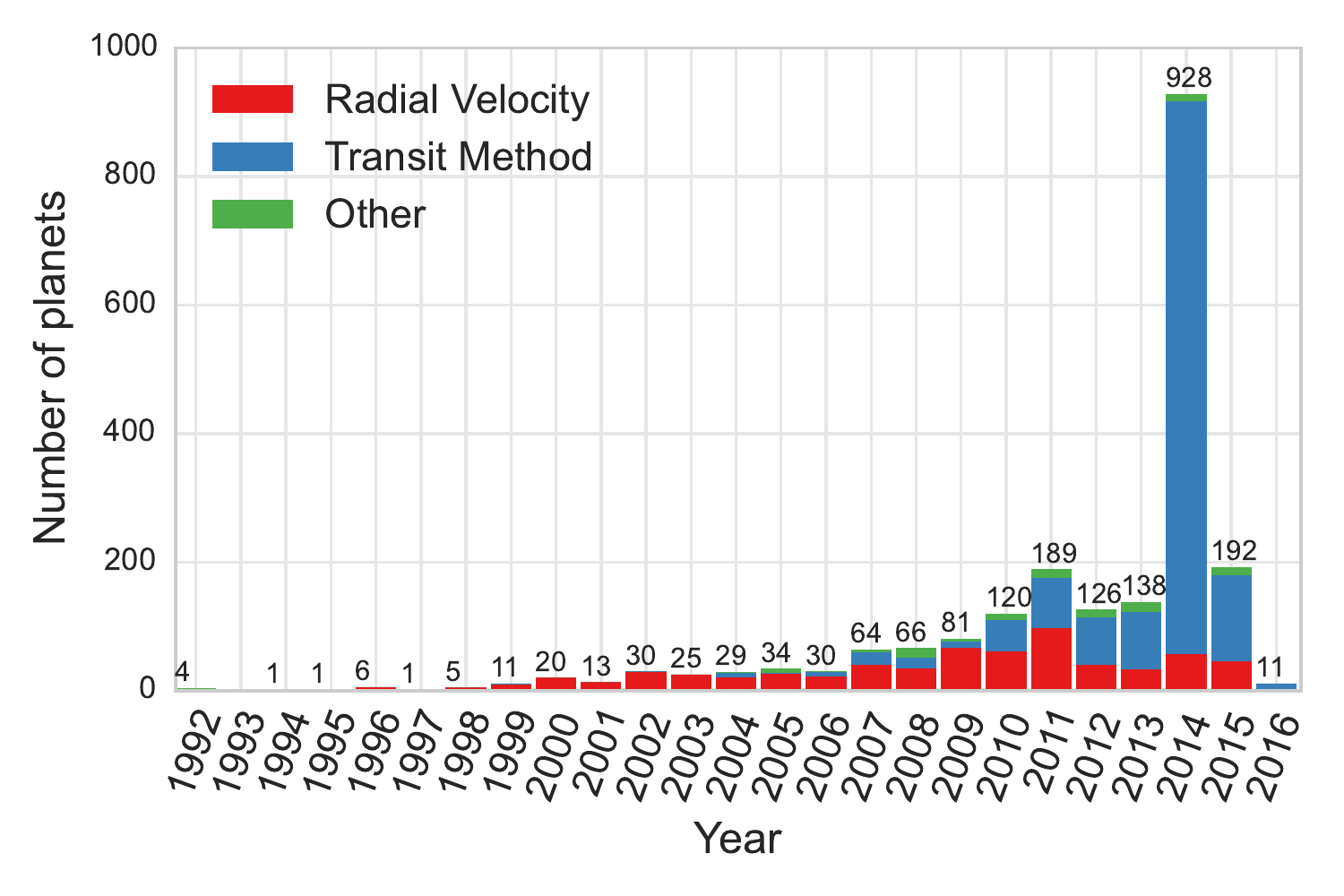}
\label{fig:exodata-dm}
\end{figure}

This plots the number of planets discovered by each type of discovery method per year as a stacked bar chart. The format is
\begin{verbatim}
exodata.plots.DiscoveryMethodByYear(planet_list, methods_to_plot
  ).plot(method_labels)
\end{verbatim}

\noindent where \textit{planet\_list} is a list of planet objects to plot, \textit{methods\_to\_plot} is the discovery methods to include (in OEC syntax) and \textit{method\_labels} are the labels to use in the legend for each discovery method. We can create a plot of the radial velocity and transit discovery methods with the following (shown in Fig. \ref{fig:exodata-dm}).

\begin{verbatim}
exodata.plots.DiscoveryMethodByYear(
    exocat.planets,
    methods_to_plot=("RV", "transit", "Other")
).plot(
    method_labels=("Radial Velocity", "Transit Method", "Other"))
\end{verbatim}

By specifying `Other' as a discovery method ExoData will group all methods not specified into this category on the plot.

\subsubsection{Plotting a parameter against another}

\begin{figure}[htbp]
  \caption{Example plots that can be generated with \textit{GeneralPlotter} in the \exodata{ }plotting module. See text for descriptions.}
  \centering
    \begin{tabular}{cc}
    \includegraphics[width=0.5\textwidth]{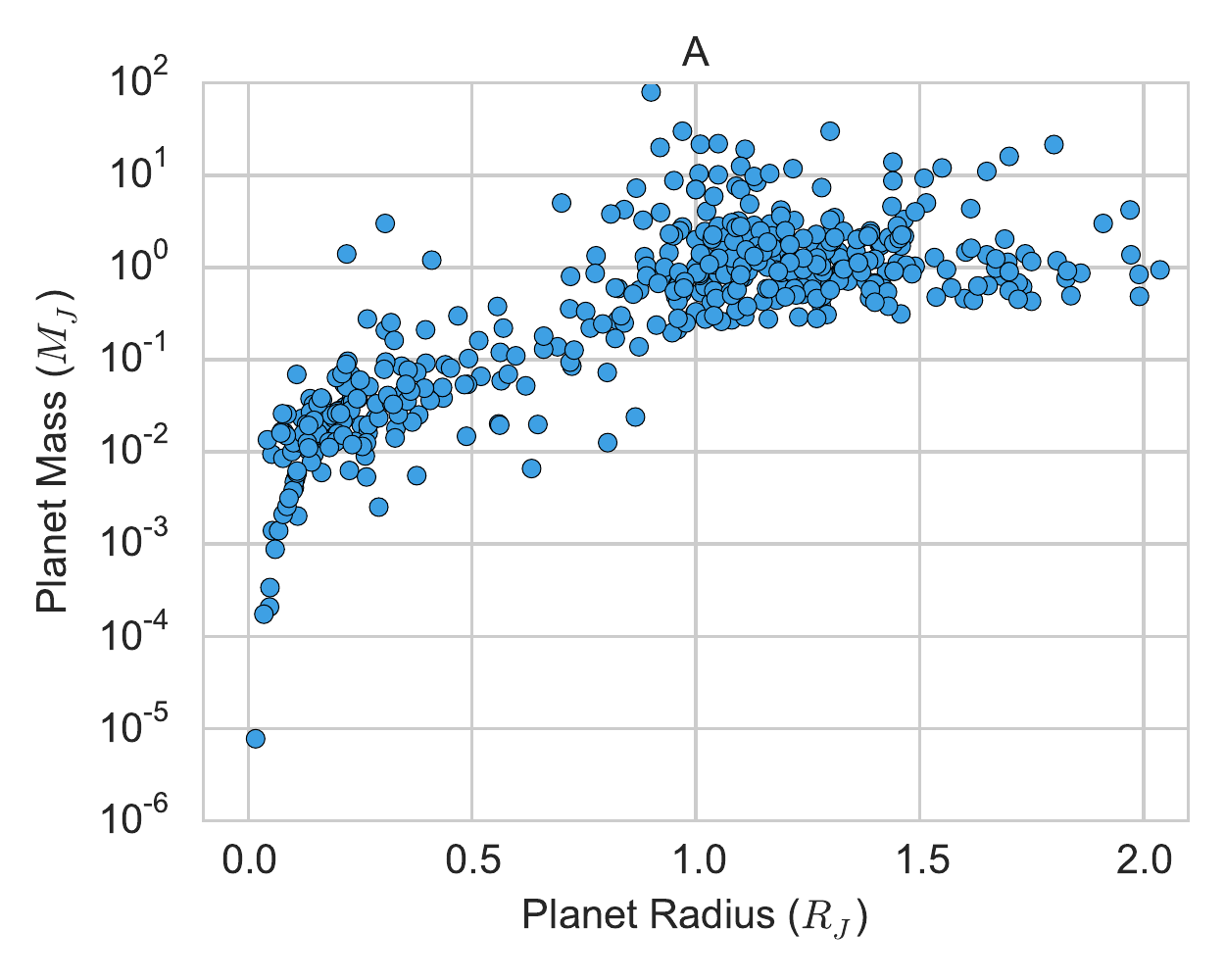}&
    \includegraphics[width=0.5\textwidth]{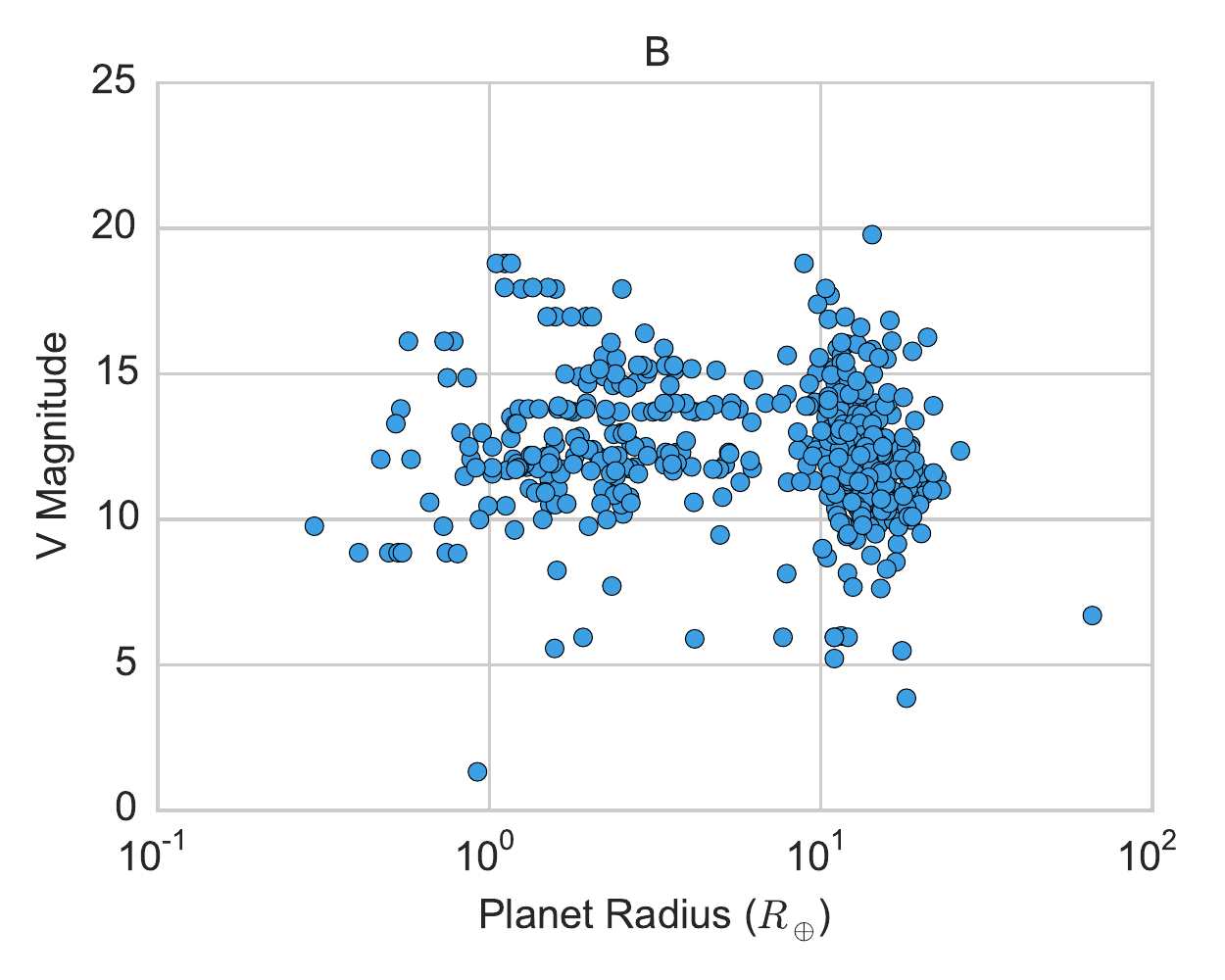}\\
    \includegraphics[width=0.5\textwidth]{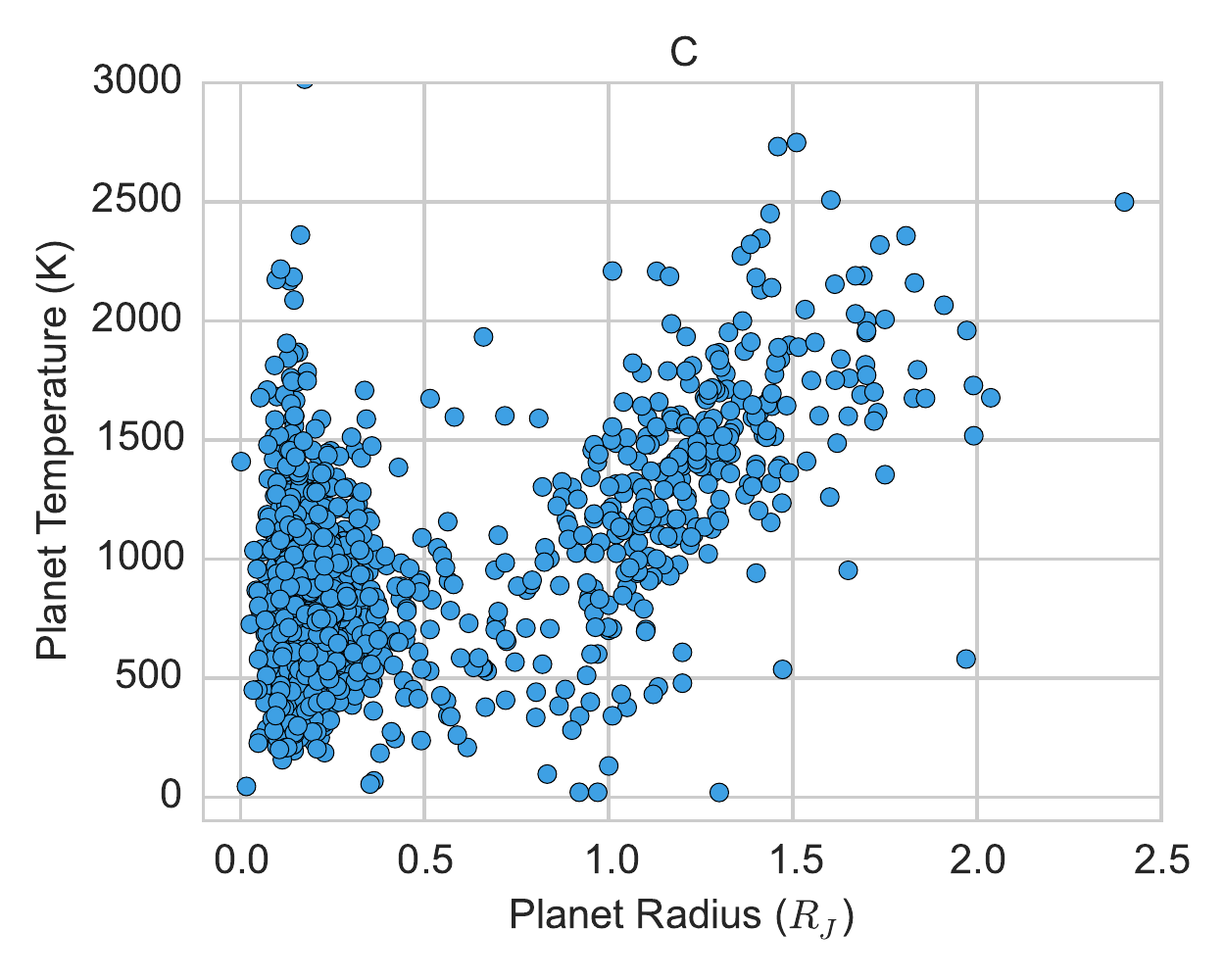}&
    \includegraphics[width=0.5\textwidth]{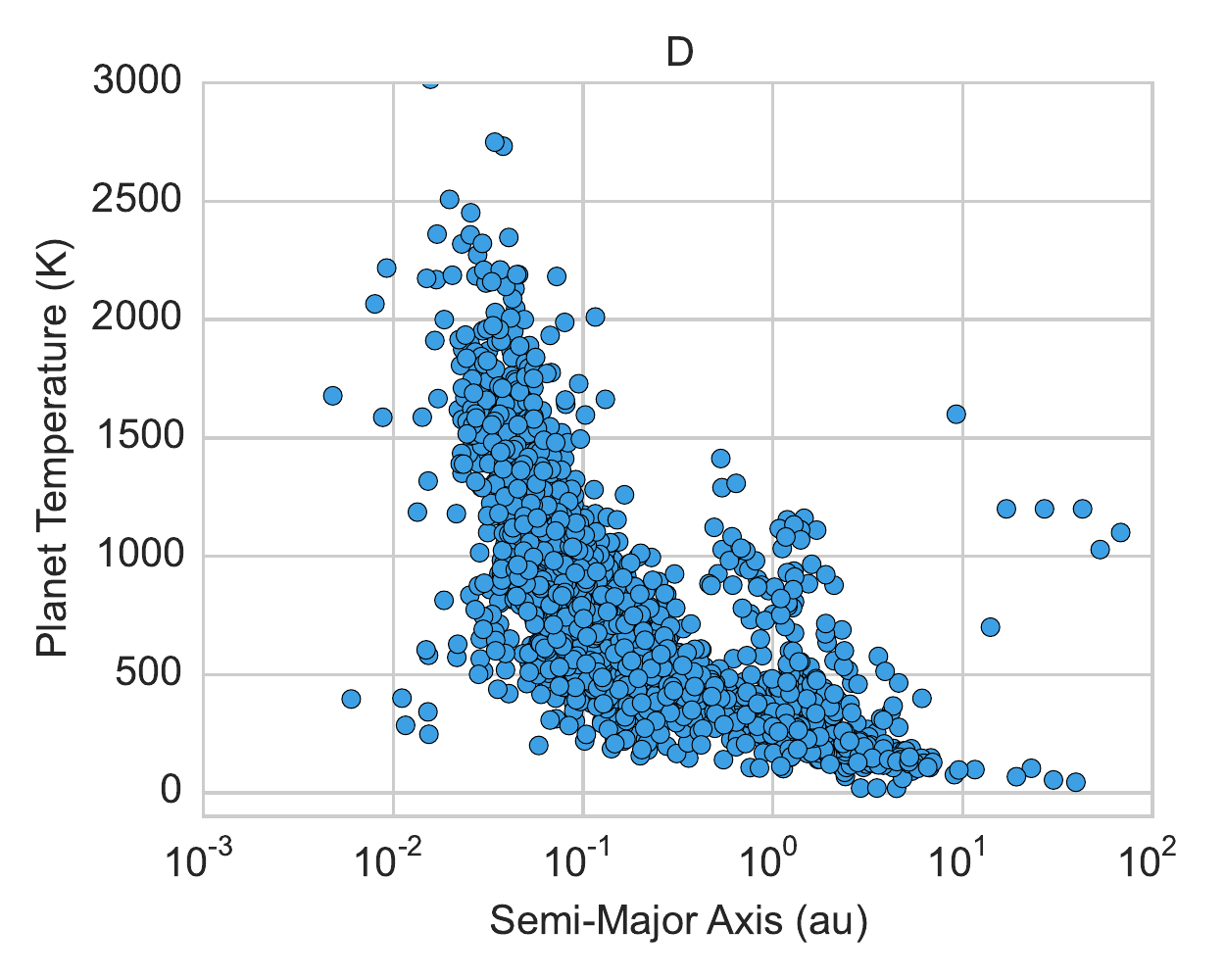}\\
    \end{tabular}
\label{fig:exodata-generalplotter}
\end{figure}

This type of plot takes a measured or calculated parameter of the planet, star, binary or system and displays it against another. The format is
\begin{verbatim}
exodata.plots.GeneralPlotter(list_of_objects,
  x_param, y_param).plot(*plot_arguments)
\end{verbatim}

\noindent where \textit{list\_of\_objects} is a list of planets, stars or binaries you want to be plotted and the x and y parameters refer to the string of the parameter you want. For example, to plot planet mass against radius (Fig. \ref{fig:exodata-generalplotter}a)
\begin{verbatim}
exodata.plots.GeneralPlotter(
    exocat.planets, "R", "M", yaxislog=True).plot()
plt.xlim(-0.1, 2.1)
\end{verbatim}

\noindent where \textit{R} would retrieve the planetary radius (\textit{planet.R}) and \textit{M} the planetary mass. Extra options are available to change the plot including log scale, changing the unit used (conversion is performed on the fly), marker colour and size and using parameters from a related object such as the parent star. Some more examples are shown in Fig. \ref{fig:exodata-generalplotter}.

\subsubsection{Number of planets per parameter bin}

These plots take input of the object list, the parameter to plot and the bin limits to produce a histogram-like plot or pie chart. The format is

\begin{verbatim}
exodata.plots.DataPerParameterBin(
    list_of_objects, parameter, bin_limit
    ).plotBarChart(*plot_arguments)
\end{verbatim}

\begin{figure}[htbp]
  \caption{Example plots that can be generated with \textit{DataPerParameterBin} in the \exodata{ }plotting module. See text for description.}
  \centering
    \begin{tabular}{cc}
    \includegraphics[width=0.5\textwidth]{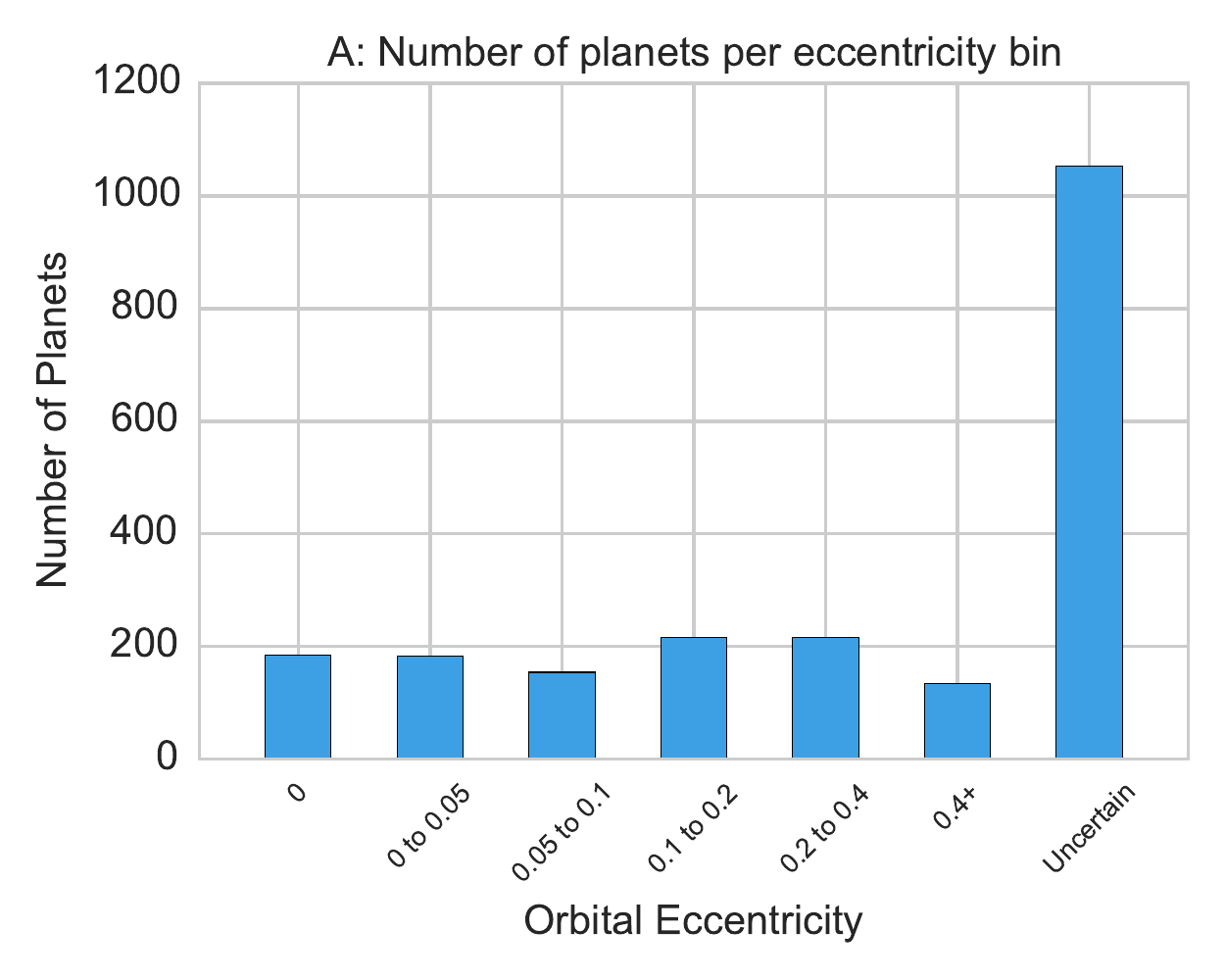}&
    \includegraphics[width=0.5\textwidth]{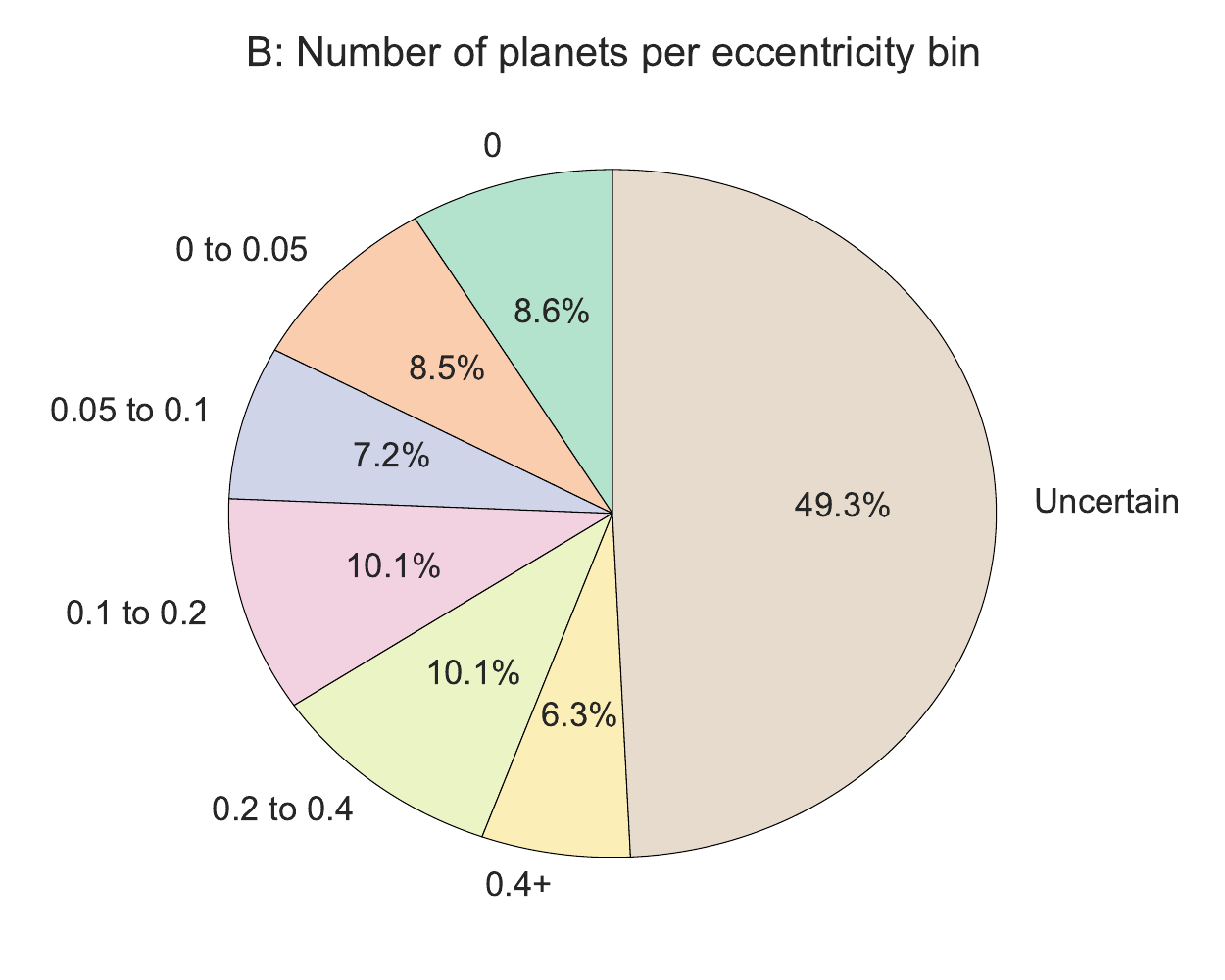}\\
    \end{tabular}
\label{fig:exodata-parameterbin}
\end{figure}

Less than / greater than limits can be made by setting the first or last limit to \textit{float(`-inf')} or \textit{float(`inf')} i.e. the following generates Fig. \ref{fig:exodata-parameterbin}A.

\begin{verbatim}
exodata.plots.DataPerParameterBin(
    exocat.planets, "e",
    (0, 0, 0.05, 0.1, 0.2, 0.4, float("inf"))
    ).plotBarChart(label_rotation=45)
\end{verbatim}

You can alternatively plot bar charts by typing \textit{.plotPieChart()} instead of \textit{.plotBarChart()} (see Fig. \ref{fig:exodata-parameterbin}B).

\section{Conclusion}

\exodata{ }provides a rounded tool for researchers wishing to use parameters of an exoplanet or exoplanetary system in code. Uses of the module include querying the exoplanet database for targets for observing proposals, running planets through a simulator and easy access to the catalogue for population statistics.

By using a version controlled catalogue like the Open Exoplanet Catalogue and the \exodata{ }interface researchers can more easily verify published results by easily obtaining the exact same planet and system parameters used by quoting both the catalogue and interface version. This is different from most catalogues which only show the latest version of which many unknown values may have changed. The interface builds upon the catalogue by making it easier to access parameters, calculate values and query the database.

\section{Acknowledgements}
This work is supported by a UCL IMPACT studentship and STFC grant code ST/P000282/1.

\bibliographystyle{elsarticle-harv}
\bibliography{exodatapaper}

\pagebreak
\appendix

\section{List of equations}
\centering
\label{sec:equation-table}
\begin{table}[hp]
	\caption{List of equations in the equations module}
    \begin{tabular}{lp{4cm}p{4cm}}
    Function name             & Function Arguments                   & Description \\
    \hline
    ScaleHeight               & T\_eff, mu, g, H                     & Eqn. \ref{eqn:scale-height}  \\
    MeanPlanetTemp            & A, T\_s, R\_s, a, epsilon, T\_p      & Eqn. \ref{eqn:greenhouse}          \\
    StellarLuminosity         & R, T, L                              & Eqn. \ref{eqn:luminosity}          \\
    KeplersThirdLaw           & a, M\_s, P, M\_p                     & Eqn. \ref{eqn:keplersthird}          \\
    SurfaceGravity            & M, R, g                              & Eqn. \ref{eqn:surface-grav}          \\
    Logg                      & M, R, logg                           & base 10 log of SurfaceGravity          \\
    TransitDepth              & R\_s, R\_p, depth                     & square of the ratio of the planetary radius to the stellar radius          \\
    Density                   & M, R, Density                        & Density = Mass/Volume          \\
    TransitDuration           & P, a, Rp, Rs, i, e, w & Eqn. \ref{eqn:duration-kipping}  \\
    ImpactParameter           & a, R\_s, i, b & Eqn. \ref{eqn:impact-parameter}          \\
	transitDurationCircular   & P, R\_s, R\_p, a, i & Eqn. \ref{eqn:duration}          \\
    estimateStellarTemperature& M\_s & Eqn. \ref{eqn:stellar-t}          \\
    estimateStellarRadius     & M\_s & Eqn. \ref{eqn:stellar-r}          \\
    estimateDistance          & m, M, Av & Eqn. \ref{eqn:star-distance}          \\
    estimateAbsoluteMagnitude & spectralType & Eqn. \ref{eqn:star-distance}          \\
    \end{tabular}
    \label{tab:equations}
\end{table}

\pagebreak
\section{List of methods provided per object in the astroclasses module}
\label{sec:methods-per-object}

The following tables (Table \ref{tab:methods-system} to \ref{tab:methods-planet}) list the methods (functions) and attributes (variables) accessible from each object. Note that objects are nested and so while the planet object has a `ra' method this is actually passed to the system object.

\begin{table}[hp]
\centering
\caption{List of System methods and attributes}
    \begin{tabular}{lp{6cm}l}
    Name                     & Description                                         & Method / Attribute \\
    \hline
    Right Ascension          & ~                                                   & .ra                \\
    Declination              & ~                                                   & .dec               \\
    Distance                 & To the system (parsecs)                             & .d                 \\
    Child Stars and Binaries & List of stars and binaries within the system & .stars             \\
    Epoch & Epoch for the orbital elements (BJD) & .epoch \\
    Name & Name of the system & .name \\
    Child objects & All objects that are children of the system, i.e. stars and/or binaries & .children \\
    Flags & Flags object (see section \ref{sec:flags}) & .flags \\
    Parameters & Dictionary holding all parameters loaded from the Open Exoplanet Catalogue & .params \\
    \end{tabular}
    \label{tab:methods-system}
\end{table}

\begin{table}[hp]
\centering
\caption{List of Binary methods and attributes}
    \begin{tabular}{lp{6cm}l}
    Name                     & Description                                         & Method / Attribute \\
    Right Ascension          & ~                                                   & .ra                \\
    Declination              & ~                                                   & .dec               \\
    Distance                 & To the system (parsecs)                             & .d                 \\
    Name & Name of the binary & .name \\
    Child objects & All objects that are children of the binary, i.e. stars, binaries and planets & .children \\
    Parent object & Object above this one, i.e a system or a binary & .parent \\
    Flags & Flags object (see section \ref{sec:flags}) & .flags \\
    Parameters & Dictionary holding all parameters loaded from the Open Exoplanet Catalogue & .params \\
    System & The system this object is in & .system \\
    Child Stars and Binaries & List of stars and binaries within the Binary (list) & .stars             \\
 Inclination & Orbital inclination & .i \\
 Eccentricity & Orbital eccentricity & .e \\
 Period & Orbital period & .P \\
 Semi-Major Axis & Orbital semi-major axis & .a \\
 Separation & The projected separation from one component to the other & .separation \\
 Transit Time & Time of the mid-point of a transit (BJD) & .transittime \\
 Periastron & Longitude of periastron (degree) & .periastron \\
 Longitude & Mean longitude at a given Epoch (degree) & .longitude \\
 Ascending node & Longitude of the ascending node (degree) & .ascendingnode \\
    U Magnitude                            & ~                                                                                                                               & .magU                    \\
    B Magnitude                            & ~                                                                                                                               & .magB                    \\
    H Magnitude                            & ~                                                                                                                               & .magH                    \\
    I Magnitude                            & ~                                                                                                                               & .magI                    \\
    J Magnitude                            & ~                                                                                                                               & .magJ                    \\
    K Magnitude                            & ~                                                                                                                               & .magK                    \\
    V Magnitude                            & ~                                                                                                                               & .magV                    \\
    L Magnitude                            & ~                                                                                                                               & .magL                    \\
    M Magnitude                            & ~                                                                                                                               & .magM                    \\
    N Magnitude                            & ~                                                                                                                               & .magN

    \end{tabular}
    \label{tab:methods-binary}
\end{table}

\begin{table}[hp]
	\centering
\caption{List of Star methods and attributes}
    \begin{tabular}{p{5cm}p{6cm}l}
    Name                                   & Description                                                                                                                     & Method / Attribute       \\ \hline
    Right Ascension                        & ~                                                                                                                               & .ra                      \\
    Declination                            & ~                                                                                                                               & .dec                     \\
    Distance                               & To the system (parsecs)                                                                                                         & .d                       \\
    Name & Name of the star & .name \\
    Child objects & All objects that are children of the star, i.e planets & .children \\
    Parent object & Object above this one, i.e a system or a binary & .parent \\
    Flags & Flags object (see section \ref{sec:flags}) & .flags \\
    Parameters & Dictionary holding all parameters loaded from the Open Exoplanet Catalogue & .params \\
    System & The system this object is in & .system \\
    Child Planets                          & List of orbiting planets & .stars                   \\
    Radius                                 & ~                                                                                                                               & .R                       \\
    Mass                                   & ~                                                                                                                               & .M                       \\
    Temperature                            & Temperature of the star in the catalogue. If absent calculates using .calcTemperature() and sets `Calculated Temperature' flag. & .T                       \\
 Age & Age of the planet (Gyr) & .age \\
    Calculate Temperature                  & Uses equations.estimateStellarTemperature                                   & .calcTemperature()       \\
    Calculate Surface Gravity              & Uses equations.SurfaceGravity                                        & .calcSurfaceGravity()    \\
    Calculate log(g)                       & Uses equations.Logg                                                   & .calcLogg()              \\
    Metalicity                             & ~                                                                                                                               & .Z                       \\
    U Magnitude                            & ~                                                                                                                               & .magU                    \\
    B Magnitude                            & ~                                                                                                                               & .magB                    \\
    H Magnitude                            & ~                                                                                                                               & .magH                    \\
    I Magnitude                            & ~                                                                                                                               & .magI                    \\
    J Magnitude                            & ~                                                                                                                               & .magJ                    \\
    K Magnitude                            & ~                                                                                                                               & .magK                    \\
    V Magnitude                            & ~                                                                                                                               & .magV                    \\
    L Magnitude                            & ~                                                                                                                               & .magL                    \\
    M Magnitude                            & ~                                                                                                                               & .magM                    \\
    N Magnitude                            & ~                                                                                                                               & .magN                    \\
    Spectral Type                          & ~                                                                                                                               & .spectralType            \\
    Calculate Density                      & Uses equations.Density to estimate density from mass and radius                                                                 & .calcDensity()           \\
    Calculate Luminosity                   & Uses equations.StellarLuminosity to estimate stellar luminosity from R and T                                                       & .calcLuminosity()        \\
    Estimate Absolute Magnitude & Uses equations.estimateAbsoluteMagnitude and the spectral type to estimate the stars absolute magnitude & .estimateAbsoluteMagnitude() \\
    Estimate Distance & Uses equations.estimateDistance to estimate the distance from Earth to the star from the absolute magnitude and V magnitude & .estimateDistance() \\
    \end{tabular}
    \label{tab:methods-star}
\end{table}

\begin{table}[hp]
	\centering
\caption{List of Planet methods and attributes}
    \begin{tabular}{p{5cm}p{6cm}l}
 Name & Description & Method / Attribute \\ \hline
 Right Ascension & ~ & .ra \\
 Declination & ~ & .dec \\
 Distance & To the system (parsecs) & .d \\
    Name & Name of the planet & .name \\
    Parent object & Object above this one, i.e binary or star & .parent \\
    Flags & Flags object (see section \ref{sec:flags}) & .flags \\
 Parent Star & Host star of the planet (if present) & .star \\
 Parameters & Dictionary holding all parameters loaded from the Open Exoplanet Catalogue & .params \\
    System & The system this object is in & .system \\
 Radius & ~ & .R \\
 Mass & ~ & .M \\
 Inclination & Orbital Inclination & .i \\
 Eccentricity & ~ & .e \\
 Period & Orbital Period & .P \\
 Semi-Major Axis & Semi-Major Axis & .a \\
 Seperation & The projected separation of planet from its host & .seperation \\
 Age & Age of the planet (Gyr) & .age \\
 Transit Time & Time of mid-point of a transit (BJD) & .transittime \\
 Periastron & Longitude of periastron (degree) & .periastron \\
 Longitude & Mean longitude at a given Epoch (degree) & .longitude \\
 Ascending node & Longitude of the ascending node (degree) & .ascendingnode \\

 Discovery Method & ~ & .discoveryMethod \\
 Discovery Year & ~ & .discoveryYear \\
 Description & Open Exoplanet Catalogue description of this planet & .description \\

 Temperature & Temperature of the star in the catalogue. If absent calculates using .calcTemperature() and sets 'Calculated Temperature' flag. & .T \\
 Calculate Temperature & Uses equations.MeanPlanetTemp & .calcTemperature() \\
 Calculate Surface Gravity & Uses equations.SurfaceGravity & .calcSurfaceGravity() \\
 Calculate log(g) & Uses equations.logg & .calcLogg() \\
 Calculate Density & Uses equations.Density & .calcDensity() \\
 Calculate Period & Uses equations.calcPeriod & .calcPeriod() \\
 Calculate Semi-Major axis & uses equations.KeplersThirdLaw & .calcSMA() \\
 Calculate Semi-Major axis from Planet temperature & Uses equations.MeanPlanetTemp  & .calcSMAfromT() \\

 Transiting Planet? & Returns true if the planet transits & .isTransiting \\
 Transit Duration & Uses equations.TransitDuration or transitDurationCircular if circular=True & .calcTransitDuration() \\
 Transit Depth & Uses equations.TransitDepth & .calcTransitDepth() \\
 Planet Type & The type of planet (ie super-Earth) as defined in assumptions. Uses assumptions.planetType which prioritises mass type & .type() \\
\end{tabular}
\label{tab:methods-planet}
\end{table}

\pagebreak
\section{List of additional units}

\begin{table}[hp]
	\centering
	\caption{List of astronomical units}
    \begin{tabular}{lll}
    Unit               & Value                     & Description            \\ \hline
    L\_s               & $3.839 \times 10^{26}$ W  & Solar Luminosity ($L_\odot$)           \\
    R\_s               & $6.995 \times 10^8$ m     & Solar Radius ($R_\odot$)   \\
    R\_e               & $6.371 \times 10^6$ m     & Earth's Radius ($R_\oplus$)\\
    R\_j               & $6.9911 \times 10^7$ m    & Jupiter's Radius ($R_j$)          \\
    M\_s               & $1.99\times 10^{30}$ kg   & Solar Mass ($M_\odot$)        \\
    M\_e               & $3.839\times 10^{26}$ kg  & Earth's Mass ($M_\oplus$)       \\
    M\_j               & $1.8986\times 10^{27}$ kg & Jupiter's Mass ($M_j$)             \\
    Gyear              & $10^9$ years              & Gigayear                   \\
    JulianDay          & 1 day                     & JD                         \\
    ModifiedJulianDate & 1 day                     & MJD                        \\
    gcm3               & $g/cm^3$                  & $g/cm^3$                   \\
    kgm3               & $kg/m^3$                  & $kg/m^3$                   \\
    ms2                & $m/s^2$                   & $m/s^2$                    \\
    \end{tabular}
    \label{tab:units}
\end{table}

\end{document}